\documentclass[12pt]{iopart}

\usepackage{graphicx}
\usepackage{bm}
\usepackage{subfig}
\usepackage{braket}
\usepackage{times}
\expandafter\let\csname equation*\endcsname\relax
\expandafter\let\csname endequation*\endcsname\relax
\usepackage{amsmath}
\usepackage{amssymb}
\usepackage{amstext}
\usepackage[sort&compress,numbers]{natbib}

\newcommand{\nat}{Nature (London)}
\newcommand{\ofr}{(\mathbf{r})}

\newcommand{\nR}{{n_\mathrm{R}}}
\newcommand{\nL}{{n_\mathrm{L}}}
\newcommand{\hnR}{{\hat{n}_\mathrm{R}}}
\newcommand{\hnL}{{\hat{n}_\mathrm{L}}}
\newcommand{\R}{\mathrm{R}}
\renewcommand{\L}{\mathrm{L}}
\newcommand{\MO}{\mathrm{MO}}

\newcommand{\BC}{\mathrm{BC}}
\newcommand{\BCMO}{{\mathrm{BC,MO}}}
\newcommand{\0}{{(0)}}
\renewcommand{\a}{{(\alpha)}}
\renewcommand{\b}{{(\beta)}}
\newcommand{\abcd}{{(\alpha\beta\gamma\delta)}}
\newcommand{\tot}{{\mathrm{tot}}}
\newcommand{\eff}{{\mathrm{eff}}}
\newcommand{\ER}{E_\mathrm{R}}
\newcommand{\as}{a_\mathrm{s}}
\newcommand{\expect}[1]{\langle #1 \rangle}
\newcommand{\I}{\mathrm{I}}
\newcommand{\F}{\mathrm{F}}
\newcommand{\hb}[2]{\hat{b}_\text{#1}^{(#2)}}
\newcommand{\hbd}[2]{\hat{b}_\text{#1}^{(#2)\dagger}}
\newcommand{\dr}{\int\! d^3{r}\ }

\newcommand{\hatt}{\tilde}
\newcommand{\Fig}[1]{Fig.~\ref{#1}}
\newcommand{\gt}{\tilde{g}}  

\begin{document}

\title[Multi-orbital and density-induced tunneling of bosons in optical lattices]{Multi-orbital and density-induced tunneling of bosons \\in optical lattices} 
\author{Dirk-S\"oren L\"uhmann, Ole J\"urgensen, and Klaus Sengstock}
\address{Institut f\"ur Laser-Physik, Universit\"at Hamburg, Luruper Chaussee 149, 22761 Hamburg, Germany}


\begin{abstract}
We show that multi-orbital and density-induced tunneling have a significant impact on the phase diagram of bosonic atoms in optical lattices. Off-site interactions lead to density-induced hopping, the so-called bond-charge interactions, which can be identified with an effective tunneling potential and can reach the same order of magnitude as conventional tunneling. In addition, interaction-induced higher-band processes also give rise to strongly modified tunneling, on-site and bond-charge interactions. We derive an extended occupation-dependent Hubbard model with multi-orbitally renormalized processes and compute the corresponding phase diagram. It substantially deviates from the single-band Bose-Hubbard model and predicts strong changes of the superfluid to Mott-insulator transition. In general, the presented beyond-Hubbard physics plays an essential role in bosonic lattice systems and has an observable influence on experiments with tunable interactions.
\end{abstract}


\maketitle


\section{Introduction}

Hubbard models are extremely successful in describing a variety of systems ranging from electrons in solids to bosonic quantum gases in optical lattices. While in solids all kinds of complex lattice systems are realized, optical lattices offer simple geometries in combination with precisely controllable experimental conditions.
The striking idea that ultracold bosonic atoms in optical lattices constitute an excellent tool for studying the superfluid to Mott-insulator transition \cite{Jaksch1998}  has initiated an intensive investigation of bosonic lattice models. It has caused a  large number of  theoretical studies on the Bose-Hubbard model ranging from mean-field treatments \cite{Fisher1989,Krauth1992,Oosten2001} and quantum Monte Carlo \cite{CapogrossoSansone2007} to DMRG \cite{Kuhner2000} approaches. In addition, the influence of disorder \cite{Fisher1989,Scalettar1991,Sheshadri1995,Damski2003,Muth2008,Pisarski2011} and the effect of confining potentials have been discussed \cite{Kashurnikov2002,Batrouni2002,Wessel2004,DeMarco2005}.
 However, the first-order corrections to the Bose-Hubbard model itself, namely higher orbital tunneling (Fig.~1a) and bond-charge interactions (Fig.~1b), have so far been mostly neglected in these approaches. 
Bond-charge interactions known from solids \cite{Hirsch1989,Strack1993,Hirsch1994,Amadon1996} constitute a density-induced tunneling process caused by the scattering of particles on neighboring sites. This extension to the Hubbard model was discussed in the context of superconductivity \cite{Hirsch1989} and ferromagnetism \cite{Amadon1996}. 
 While for fermionic systems these effects are partly suppressed due to the Pauli principle, they have large impact on bosonic systems. 
The unique experimental access in optical lattices allows for a detailed
investigation of this interaction-assisted tunneling.
This density-dependent hopping was recently also discussed for boson-fermion mixtures \cite{Mering2010,Luhmann2008b}, where
 the Bose-Fermi-Hubbard model does not fully cover effects arising from the interspecies interaction \cite {Luhmann2008b,Lutchyn2009,Best2009,Mering2010,Heinze2011}.
In this context, the role of higher orbitals for the tunneling processes \cite{Luhmann2008b,Lutchyn2009,Mering2010} was investigated. 
For bosonic systems, modifications of tunneling and on-site interaction due to higher bands were studied by variational mean-field methods \cite{Li2006,Larson2009,Hazzard2010,Dutta2011} and by numerical exact methods restricted either to double- or triple-well systems \cite{Schneider2009,Sakmann2009,Sakmann2010,Cao2011} or to the on-site interaction \cite{Busch1998, Johnson2009, Will2010, Buchler2010}. However, in the correlated regime present at the vicinity of the Mott-insulator transition the applicability of mean-field methods to describe higher-band processes is doubtful. Thus, the challenging problem is to find a comprehensive description including both bond-charge hopping \cite{Hirsch1989,Strack1993,Hirsch1994,Amadon1996,Mazzarella2006,Mering2010,Cataldo2011} and higher band processes \cite{Li2006,Larson2009,Hazzard2010,Dutta2011,Busch1998,Johnson2009,Sakmann2009,Sakmann2010, Will2010, Buchler2010, Pilati2011, Inoue2011} that is valid also for strongly correlated systems. The recent experimental progress in accessing the occupation-dependent on-site interactions \cite{Campbell2006,Will2010,Mark2011,Bakr2011} and the tunneling matrix element \cite{Heinze2011} requests for an accurate theoretical method to calculate these parameters. Furthermore, the superfluid to Mott-insulator transition in experiments with tunable interactions \cite{Best2009,Mark2011} is directly affected by bond-charge tunneling as well as multi-orbital extensions to conventional and bond-charge tunneling.


\begin{figure}
\begin{center}
\includegraphics[width=0.7\linewidth]{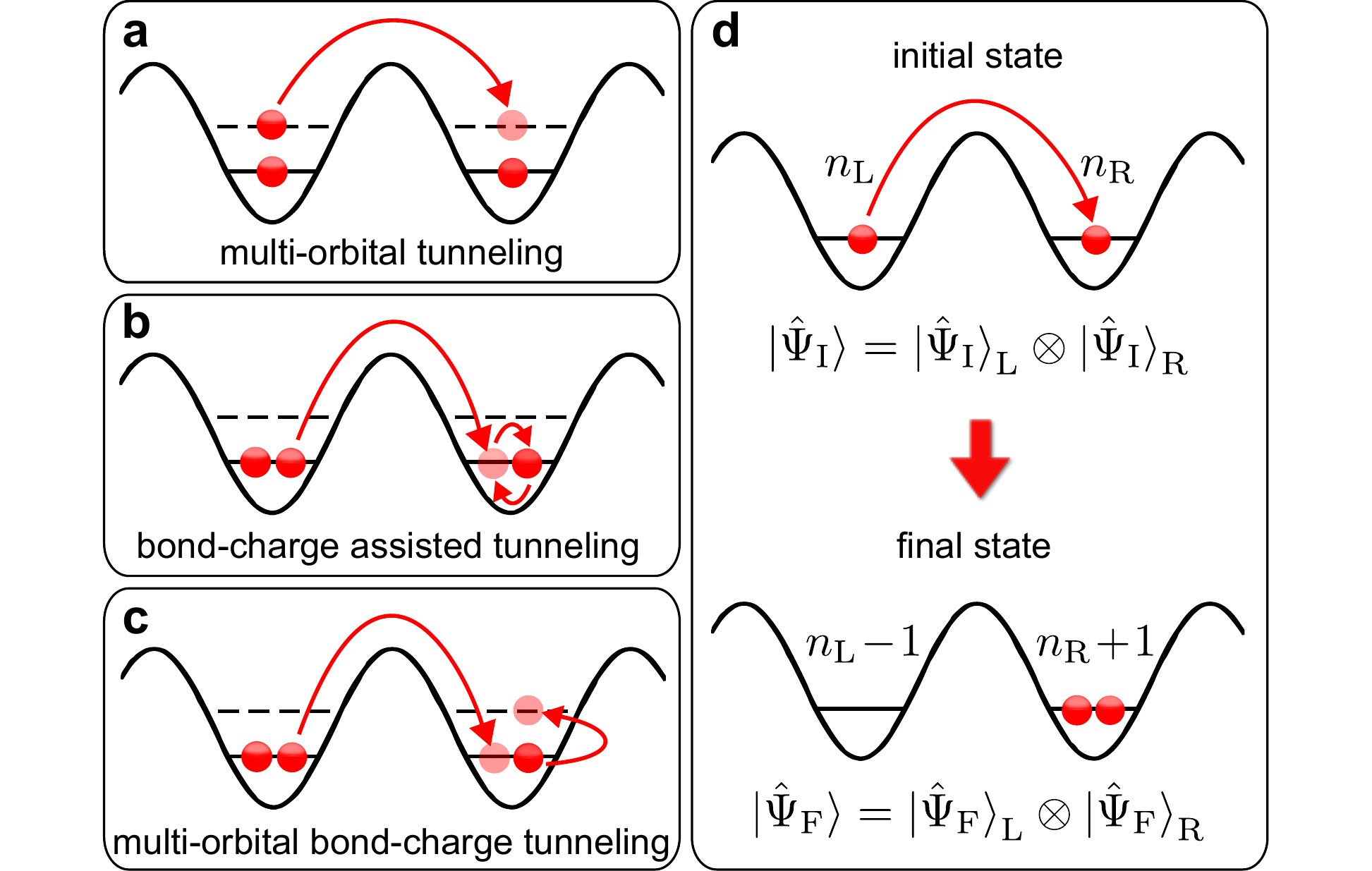}
\end{center}
\vspace{-1em}
\caption{Hopping processes beyond the standard single-band Hubbard tunneling: Example processes for (\textbf{a}) multi-orbital tunneling, (\textbf{b}) single-orbital bond-charge assisted hopping and (\textbf{c}) multi-orbital bond-charge hopping. (\textbf{d}) Initial and final states for one particle tunneling from the left to the right site. The interaction causes an admixture of higher orbital states requiring a renormalization of the bare lowest-band tunneling processes.
\label{Process}
}
\end{figure}


Here, we show that interaction-induced processes cause two substantial modifications of the Bose-Hubbard model for bosonic atoms in three-dimensional optical lattices. First, so-called bond-charge or density-induced tunneling combining interaction and hopping (Fig.~1b) is a notable contribution to tunneling. 
Second, higher bands are occupied due to the interaction of particles on a lattice site. 
This increases the tunneling significantly as the tunneling in higher bands (Fig.~1a) is strongly enhanced.
We  present a general scheme to calculate renormalized tunneling processes by effectively incorporating higher-band contributions. On this basis, we derive an effective occupation-dependent model where the renormalization solely depends on the solution of the on-site problem. The well-known phase diagram for bosons in lattices describing the superfluid (SF) to Mott insulator (MI) transition is drastically altered. The presented results have direct relevance for optical lattice experiments, which can therefore serve as an ideal testing ground for beyond Hubbard physics. 

After presenting the resulting phase diagram in section \ref{sec:PD}, we discuss the extensions of the Hubbard model individually starting with  the effect of bond-charge interactions in section \ref{BCI}.  The multi-orbital renormalization of the operators for conventional tunneling, bond-charge tunneling and on-site interaction is elaborated in sections 4-7.   
The basic idea is to separate the strongly-correlated on-site problem from the inter-site dynamics. First, we solve the on-site problem numerically exactly using the Wannier states of the lattice. Subsequently, we derive renormalized matrix elements for the multi-orbitally dressed operators.  
Finally, we calculate the phase diagram by treating the many-site problem with mean-field theory in section \ref{MFGW}. 
Note that the multi-orbital renormalization can be used for various methods to treat the many-site problem and
that the results are expected to be similarly affected.
 
\section{Phase diagram and model Hamiltonian} \label{sec:PD}
\begin{figure*}[t]
\begin{center}
\includegraphics[width=1\linewidth]{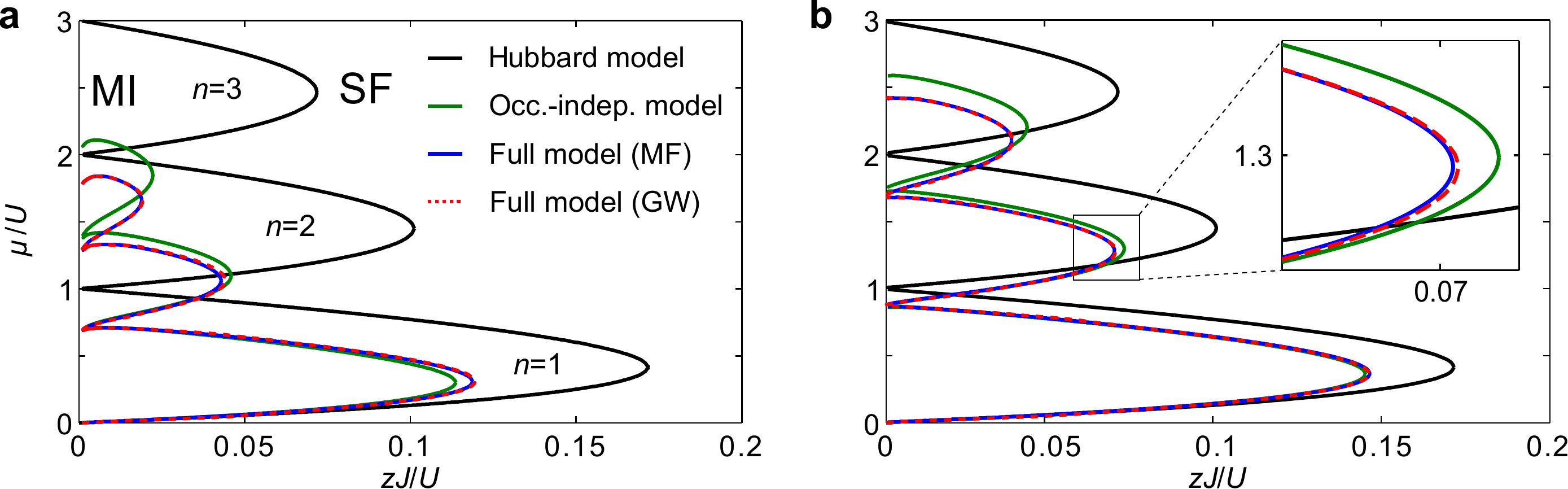}
\end{center}
\vspace{-1em}
\caption{Phase diagrams showing the superfluid to Mott-insulator transition for interaction strengths $\as$ normalized to the lattice spacing $a$ (\textbf{a}) $\as/a=0.042$ and (\textbf{b}) $\as/a=0.014$. The latter corresponds to the background scattering length $\as=100\,a_0$ of $^{87}$Rb at a lattice spacing of $a=377\,\mathrm{nm}$ \cite{Best2009}. 
The phase boundaries are plotted for the Bose-Hubbard model (black) and the full Hamiltonian Eq.~\eqref{eq:Htot} using mean-field (blue) and Gutzwiller (dashed red). The green line corresponds to the occupation-independent Hamiltonian Eq.~\eqref{eq:Hsimple}. 
\label{PD}
}
\end{figure*}

\begin{figure*}[t]
\begin{center}
\includegraphics[width=0.5\linewidth]{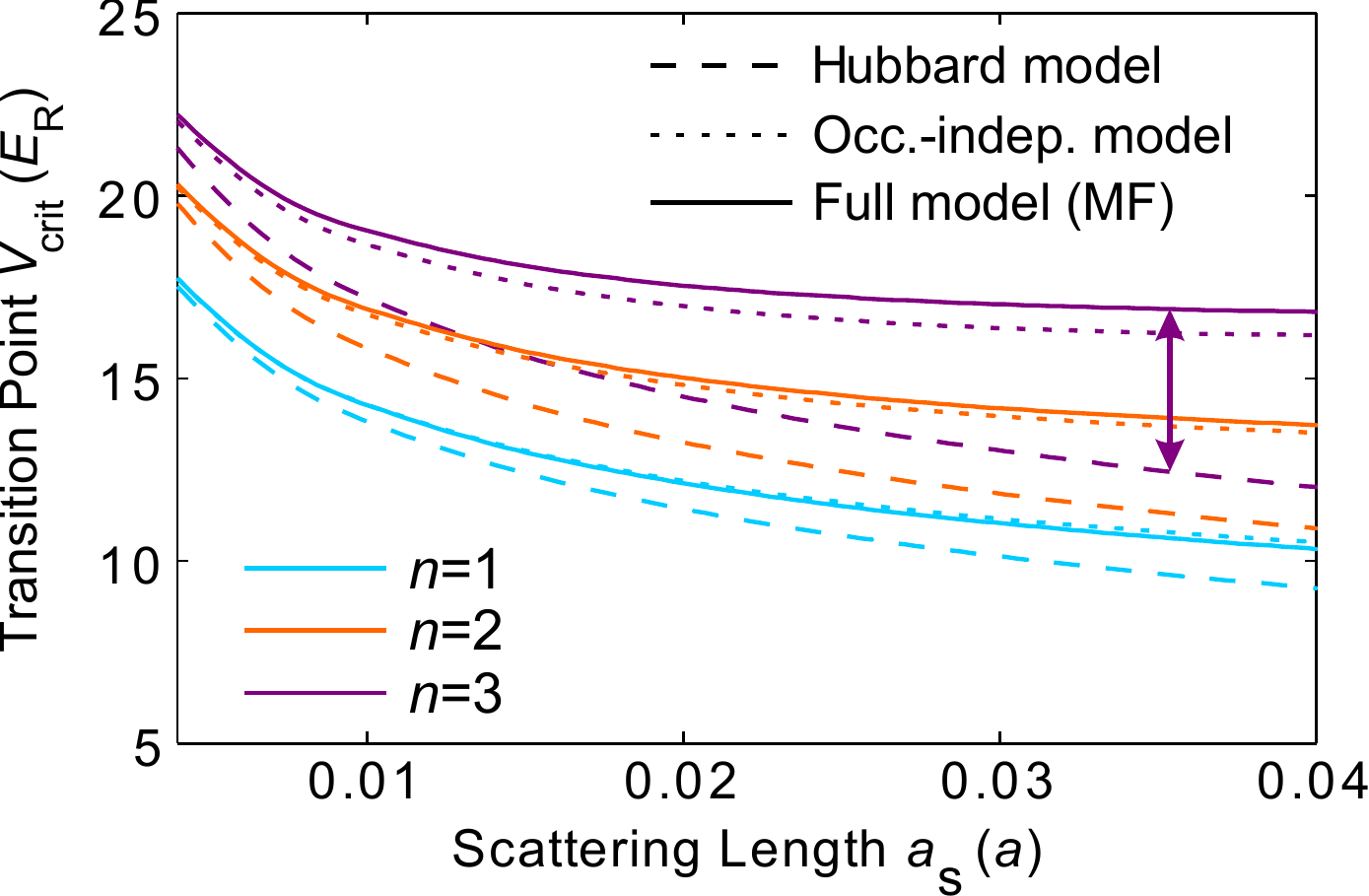}
\end{center}
\vspace{-1em}
\caption{Critical lattice depths of the  superfluid to Mott-insulator transition for the Hubbard model (dashed), the full multi-orbital Hamiltonian Eq.~\eqref{eq:Htot}  and the occupation-independent lowest-band Hamiltonian Eq.~\eqref{eq:Hsimple}. The results are plotted as a function of the scattering length $\as/a$ in units of the  recoil energy $\ER=h^2/8 m a^2$.
\label{Vc} 
}
\end{figure*}

We now first explain as one central result the phase diagram for bosonic atoms (\Fig{PD}) which is derived and further discussed afterwards. The black lines depict the Mott lobes predicted by the single-band Bose-Hubbard model applying mean-field theory \cite{Oosten2001}. Our calculations (colored lines) take into account the corrections to the tunneling as mentioned above as well as multi-orbital modifications of the on-site interaction. The phase diagrams are obtained for bosonic atoms of arbitrary species in cubic sinusoidal optical lattices with $V(\mathbf{r})=V_0\sum_i \cos^2(\pi r_i/a)$, where $i=\{x,y,z\}$, $a$ is the lattice spacing and $V_0$ the lattice depth. We predict that the phase area of the Mott insulator is considerably reduced even for moderate interaction strengths and substantially deformed for stronger interactions. This corresponds to a significant shift of the critical lattice depth of the SF-MI transition (\Fig{Vc}) due to an effectively increased tunneling and reduced on-site interaction. The discrepancy can reach more than $5\,\ER$ for $n=3$ particles per lattice site. We expect the shift to be observable in experiments with a filling $n\geq 2$ and strong or tunable interactions.
 
The mentioned extension to the Hubbard model leading to the phase diagram in \Fig{PD} (red and blue lines) can be included in an {\it effective} Hamiltonian
\begin{equation}
	\label{eq:Htot}
	\hat H_\mathrm{full}= -\sum_{\langle i,j \rangle}  \hatt{b}_i^\dagger \hatt{b}_j J^\tot_{\hatt n_j,\hatt n_i} 
	+ \frac{1}{2} \sum_i U^\MO_{\hatt n_i} \hatt n_i (\hatt n_i -1).
\end{equation}
with \textit{multi-orbitally renormalized} creation and annihilation operators $\hatt{b}^{\dagger}_i$ and $\hatt{b}_i$, respectively, and the operator $\hatt n_i=\hatt{b}^{\dagger}_i \hatt{b}_i$ corresponding to the number of particles on site $i$.	
This Hamiltonian has the same structure as the Bose-Hubbard Hamiltonian \cite{Fisher1989,Jaksch1998} but with  multi-orbitally renormalized  tunneling matrix elements  $J^\tot_{\hatt n_j,\hatt n_i}$ and on-site interactions $U^\MO_{\hatt n_i}$ that explicitly depend on the site occupation $n_i$. In the following, we calculate these occupation-dependent parameters with exact diagonalization and derive the phase boundaries using a mean-field (MF) and a Gutzwiller (GW) approach. 
Furthermore, we present a simplified Hamiltonian with \textit{occupation-independent} parameters (green lines in \Fig{PD}) which is in good agreement with the full model above. This Hamiltonian is restricted to conventional and bond-charge tunneling in the lowest-band
\begin{equation}
	\label{eq:Hsimple} \hat H_\mathrm{occ.-indep.}=
	-\sum_{\langle i,j \rangle} \Big[ J\, \hat{b}^{\dagger}_i \hat{b}_j
	+J_\BC\, \hat{b}^{\dagger}_i (\hat n_i + \hat n_j) \hat{b}_j \Big]  
	+ \frac{\tilde U_2}{2} \sum_i \hat n_i (\hat n_i-1)
\end{equation}
with single-band operators  $\hat{b}^{\dagger}_i$ and $\hat{b}_i$, the Hubbard tunneling $J$, lowest-band bond-charge interaction $J_\BC$  and renormalized on-site interaction $\tilde U_2$ for two particles, as is elaborated in detail below. In general, the results for Hamiltonian \eqref{eq:Hsimple} show that the lowest-band bond-charge tunneling is an important contribution and its influence increases with the number of particles per site. Note that the mean-field and Gutzwiller approach fully coincide for occupation-independent models (green and black lobes),
whereas there is a minimal discrepancy for the Hamiltonian \eqref{eq:Htot} with occupation-dependent tunneling.  
In the mean-field calculation, the decoupling of the lattice sites requires the use of $J^\tot_{n_i,n_i}$
as an approximation for $ J^\tot_{n_i\pm1,n_i}$ and $J^\tot_{n_i,n_i\pm1}$ (see \ref{AppendixMeanField} and G).

\section{Bond-charge interactions} \label{BCI}
\begin{figure}[t]
\begin{center}
\includegraphics[width=1\linewidth]{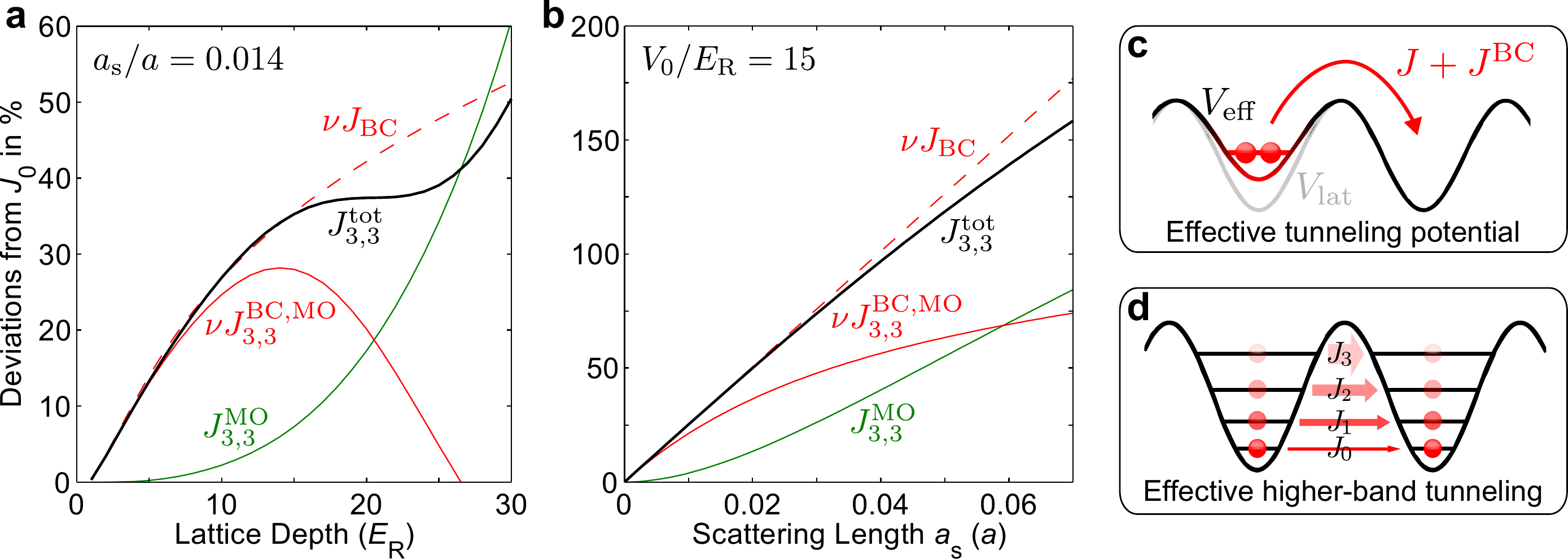}
\end{center}
\vspace{-1em}
\caption{Contributions to the effective tunneling $J^\mathrm{tot}_{n_j,n_i}$ with $n_i=n_j=3$ by multi-orbital tunneling $J^\MO_{n_j,n_i}$, bond-charge interactions $J_\BC$ and multi-orbital bond-charge interactions $J^\BCMO_{n_j,n_i}$, where the latter two scale with the prefactor $\nu=n_i+n_j-1$.
 The deviations from the bare Hubbard tunneling $J_0$ are plotted as a function of (\textbf{a}) the lattice depth $V_0$ and (\textbf{b}) the interaction strength $\as$. (\textbf{c}) The single-band bond-charge interaction gives rise to an effective tunneling potential (see text). (\textbf{d}) Qualitatively, the multi-orbital tunneling scales with the population of higher orbitals (shading) where the tunneling $J_\alpha$ (arrows) is substantially enhanced in higher bands. 
}
\label{J}
\end{figure}

The two-particle interaction of bosonic particles in the lowest band of the lattice
 is given by
   $\hat H_\mathrm{int}= \frac{1}{2} \sum_{ijkl}  U_{ijkl}\,
    \hat{b}^{\dagger}_i \hat{b}^{\dagger}_j \hat{b}_k \hat{b}_l $ 
with interaction integrals 
\begin{equation}
   \label{eq:Uijkl}
 	U_{ijkl}= g \dr 
    w^{*}_i\ofr w^{*}_j\ofr w_k\ofr w_l\ofr
\end{equation}
and Wannier functions $w_i\ofr=w^\0(\mathbf r - \mathbf r_i)$ of the lowest band, which are maximally localized at site $i$. Here, a $\delta$-interaction potential with repulsive interactions $g=\frac{4\pi\hbar}{m}\as>0$ is assumed. One of the key ingredients of the Hubbard model \cite{Hubbard1963,Fisher1989,Jaksch1998} is to restrict the two-particle interaction to the dominating term, namely, the on-site interaction $U=U_{iiii}$. 
However, some of the neglected off-site processes correspond to scattering accompanied by hopping to a neighboring site
(e.g., $ \hat{b}^{\dagger}_i \hat{b}^{\dagger}_j \hat{b}_j \hat{b}_j$) known as bond-charge tunneling (Fig.~1c). It is essential that this process constitutes a hopping process which physically modifies the overall tunneling in the system. 
Thus, although this density-dependent hopping contribution is small in comparison with the on-site interaction $U$,
it can only be neglected if also small in comparison with the conventional tunneling $J$.
Considering two neighboring sites $i$ and $j$, the density-induced tunneling operator takes the form
\begin{equation}
   \label{eq:JBC}
    \hat{J}_\BC = -J_\BC\; \hat{b}_i^\dagger (\hat{n}_i + \hat n_j) \hat{b}_j 
	\ \ \ \textrm{with} \ \ \   J_\BC=- U_\mathrm{iiij}
\end{equation}
and $J_\BC>0$. 
Due to the structure of this operator, we can combine it with the conventional tunneling $J$ to an effective hopping 
\begin{equation}
   \label{eq:JBCtot}
	\hat J^\BC_\eff=   	
	- \left[ J + J_\BC (\hat n_i + \hat n_j-1) \right] \hat{b}^{\dagger}_i \hat{b}_j.  
\end{equation}
As a consequence, 
the total tunneling increase is of the order of several ten percent for standard $^{87}$Rb-conditions (see \Fig{J}, dashed line) and even more for stronger interactions and deeper lattices. 
Note that density-density off-site interactions ($U_{ijij}$) \cite{Kuhner2000} and correlated pair tunneling ($U_{iijj}$) are considerably weaker than the bond-charge interactions (see \ref{AppendixOffSite}). 

The nature of bond-charge interactions can be intuitively illustrated as an effective potential. 
 In Eq.~\eqref{eq:JBCtot}, the term $\hat n_i + \hat n_j-1$ corresponds to the density $\rho_\mathrm{BC}(\mathbf r)= n_i |w_i\ofr|^2 + {(n_j-1)}|w_j\ofr|^2$ on sites $i$ and $j$ excluding the hopping particle. By inserting the explicit expressions for the integral $J_\BC$ and the tunneling $J$, we can write the effective hopping operator~\eqref{eq:JBCtot} as  
 \begin{equation}
	\label{eq:effectivePotential}
	\hat J^\BC_\eff=\!\int\! d^3 r\ w^{*}_i \!\left(\frac{\mathbf{p}^2}{2m}+V\ofr +g \rho_\BC \ofr \right)\! w_j\,  
	\hat{b}^{\dagger}_i \hat{b}_j.
\end{equation} 
 We can now identify the expression $V\ofr +g \rho_\BC\ofr$ as an effective tunneling potential which is illustrated in \Fig{J}c. Since the density of the repulsively interacting particles is localized at the centers of the lattice sites, the effective potential corresponds to a shallower lattice, i.e., increased tunneling. 
Note that the effective potential can be used to determine the total tunneling via band structure calculations (see \ref{AppendixEffectivePotential}). 

\section{Multi-orbital tunneling} 
While the extension to the Hubbard model above is still within the lowest-band approximation, the following discussion covers the influence of higher orbitals. The 
calculation of the effective tunneling can be performed by taking two adjacent sites of the lattices into account (Fig.~1d).
Writing initial and final states as a product of wave functions of the two sites allows to express the renormalized tunneling in terms of single-site expectation values. Consequently, the renormalized parameters depend only on the solution of the single-site problem. The presented scheme allows to derive the effective Hamiltonian \eqref{eq:Htot} independently of the applied method for treating the on-site many-particle problem. Here, the orbital occupations are computed by exact diagonalization of $n$ particles on a single site using a finite many-particle basis with a high-energy cut-off. The short-range interaction between the atoms is modeled as a box-shaped interaction potential (see \ref{AppendixModel} for details).
The single-particle orbitals are represented by Wannier functions $w^{(\alpha)}(\mathbf{r})$ with a three-dimensional band index $\alpha=(\alpha_x,\alpha_y,\alpha_z)$ .

In the following, we consider multi-orbital  hopping processes, where one atom tunnels from the left site (L) with initially $\nL$ particles to the right site (R) with initially $\nR$ particles (Fig.~1d). As a direct consequence, the total tunneling $J_ {\nL,\nR}$ becomes intrinsically occupation-dependent.
The multi-orbital hopping operator $\hat{J}^\mathrm{MO}= \sum_{\alpha} J_\alpha \,\hat{b}^{(\alpha)\dagger}_\R \hat{b}^{(\alpha)}_\L$ 
sums over all possible multi-orbital processes 
\begin{equation}
    \label{eq:multi-band-tunneling}
    J_\alpha=-\!\dr w^{(\alpha)*}_\R \left(\frac{\mathbf{p}^2}{2m}+V\ofr\right) w^{(\alpha)}_\L ,
\end{equation}
where tunneling takes place only between equal orbitals due to orthogonality relations.
It is possible to reduce this complex multi-orbital operator to an effective hopping operator
\begin{equation}
    \hat{J}^\MO_\eff=-\hatt{b}^\dagger_i \hatt{b}_j \,  J^\MO_{\hatt{n}_j,\hatt{n}_i} .
\end{equation}
The occupation-dependent tunneling matrix elements are given by  
    $J^\MO_{\nL,\nR} \propto {\bra{\Psi_\F} {\hat{J}^\MO} \ket{\Psi_\I}}$,
where $\Psi_\I=\Psi(\nL,\nR)$ denotes the initial and $\Psi_\F=\Psi(\nL-1,\nR+1)$ the final state (Fig.~1d). This expression can be separated into matrix elements for the respective lattice sites
\begin{equation}
J^\MO_{\nL,\nR} = \frac{1}{\sqrt{n_\L(n_\R+1)}} \sum_\alpha J_\alpha \bra{\Psi(\nR+1)} \hbd{}{\alpha} \ket{\Psi(\nR)}\ \bra{\Psi(\nL-1) } \hb{}{\alpha} \ket{\Psi(\nL)},
\end{equation}
where $\Psi(n)$ is the single-site wave function. This can be evaluated using the coefficients obtained by exact diagonalization as elaborated in \ref{AppendixMOT}.

The tunneling in higher bands $J_\alpha$ is drastically enhanced compared with the lowest band tunneling $J_0$. Therefore, the contributions of higher-orbital processes are significant and grow exponentially with the lattice depth, although the population of higher bands is usually smaller than 1$\%$.
As an example, the matrix element $J^\MO_{3,3}$ for three bosons per site  is plotted in \Fig{J} (green lines). In general, the effective tunneling is increased except for the elements $J^\MO_{n,0}=J^\MO_{1,n-1}$ where the tunneling is slightly reduced.

In an intuitive picture, the many-particle calculation can be reduced to the occupation of single-particle orbitals $\expect{\hat n^\a}$. The fraction of particles occupying higher orbitals $\alpha$ tunnels via the respective tunneling matrix elements $J_\alpha$, as indicated in \Fig{J}(d). 
In this approximation, the effective tunneling can be estimated by the weighted sum (see \ref{AppendixMOT})
\begin{equation}
J^\MO_{\nL,\nR} \approx \frac{1}{\sqrt{n_\L(n_\R+1)}} \sum_{\alpha} J_\alpha \sqrt{{\expect{\hat n^\a_\nL} \expect{\hat n^\a_{\nR+1}}}}.
\end{equation}  
This simplified approach is in qualitative agreement with the fully correlated calculation above. 


\section{Multi-orbital bond-charge interaction} 
It proves necessary to apply the same multi-orbital treatment of the normal tunneling developed in the last section to the bond-charge assisted hopping. 
By analogy, we can define an effective multi-orbital bond-charge hopping
\begin{equation}
\hat{J}^\BCMO_\eff =-\hatt{b}_i^\dagger (\hatt{n}_i + \hatt{n}_j)\hatt{b}_j \ J^\BCMO_{\hatt{n}_j,\hatt{n}_i}. 
\end{equation}
with an occupation-dependent parameter. It depends on the single-site matrix elements $\bra{\Psi(n-1)} \hb{}{\alpha} \ket{\Psi(n)}$ and $\bra{\Psi(n-1)}\hbd{}{\beta} \hb{}{\gamma} \hb{}{\delta}\ket{\Psi(n)}$ as discussed in \ref{AppendixBCT}.
The results shown in \Fig{J} for the multi-orbital bond-charge interaction (solid red line) differ substantially from the single-orbital case (dashed line) for intermediate and deep lattices.
Processes involving higher bands start to dominate for deep lattices, rendering the lowest-band calculation invalid, and can even turn the sign of the bond-charge contributions negative. 
Processes of the type $J^\BCMO_{n,0}$ are only weakly influenced by multi-orbital corrections.

\section{Effective tunneling}
\begin{figure}[t]
\begin{center}
\includegraphics[width=0.75\linewidth]{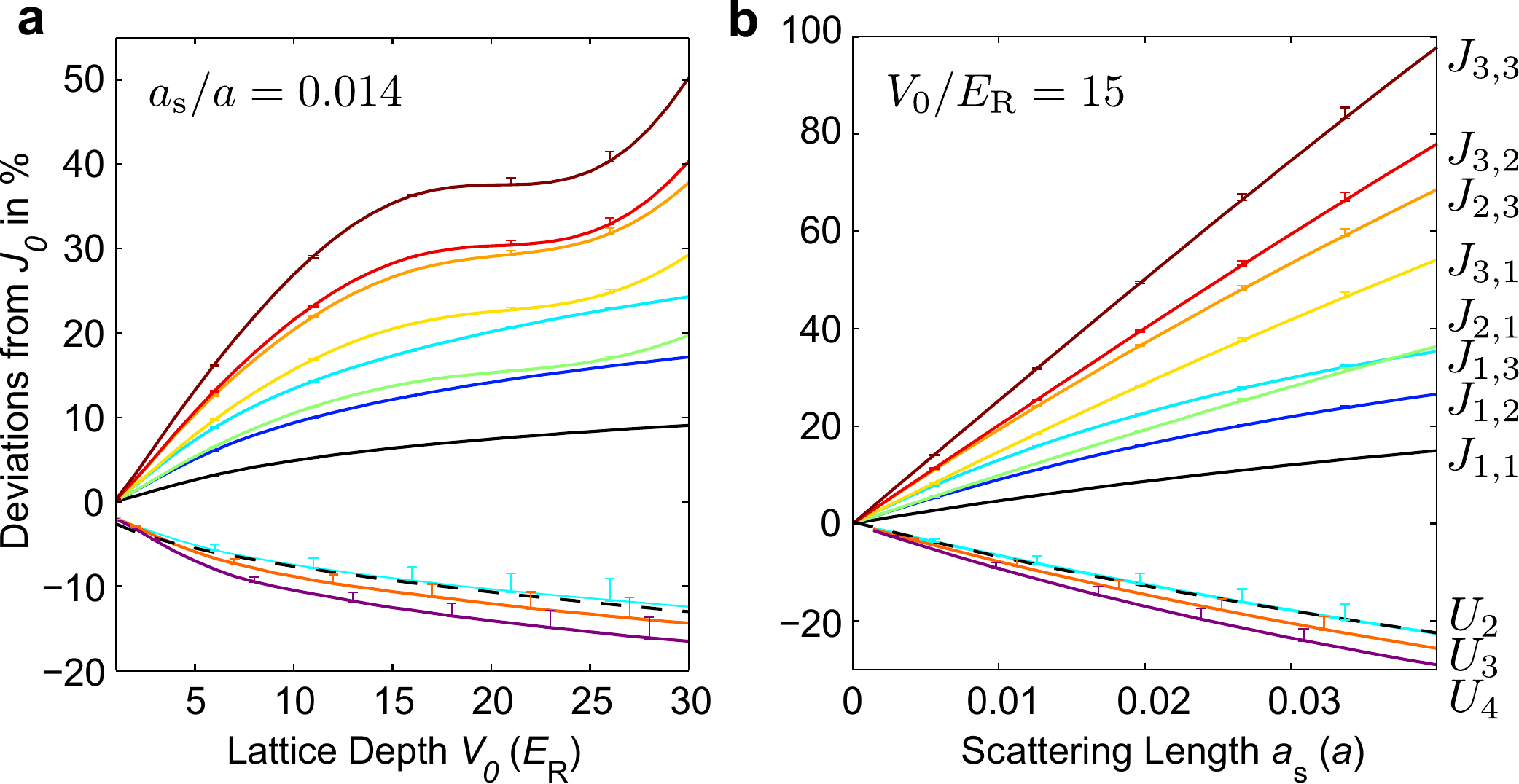}
\end{center}
\vspace{-1em}
\caption{Occupation-dependent total tunneling $J^\tot_{n_j, n_i}$ and on-site interactions $U^\MO_n$ as a function of (\textbf{a}) the lattice depth $V_0$ and (\textbf{b}) the interaction strength $\as$. The empirical fit $\tilde U_2$ is plotted as a dashed black line. Note that $J_ {n_j,n_i}$ and $J_ {n_i\!+\!1,n_j\!-\!1}$ describe time-reversed hopping processes and thus are equivalent. For an explanation of the error bars see text.
\label{Jtot}
}
\end{figure}
The total tunneling consists of normal and bond-charge assisted tunneling, both effectively including higher orbital processes. The total tunneling
\begin{equation} 
	J^\tot_{n_j,n_i}=   J^\MO_{n_j,n_i} + (n_i\! +\! n_j \! - \! 1) J^\BCMO_{n_j,n_i} 
\end{equation}
and the individual contributions are shown in \Fig{J}a and b. The occupation-dependent enhancement of the tunneling is depicted in \Fig{Jtot}. 
The deviation from the Hubbard tunneling $J$ can easily reach $30\%$ for three particles per site at moderate $\as$ and $V_0$. 
The data in this figure is calculated using a box-shaped interaction potential (see \ref{AppendixModel}) with a width $W=5\, \mathrm{nm}$ and a lattice constant $a=377\, \mathrm{nm}$. It is important to note that the results are almost independent of $W$, providing that $W$ is much smaller than the lattice constant. The error bars in \Fig{Jtot} show the results for $W=25\, \mathrm{nm}$ and $W\! \rightarrow 0$.

While the normal tunneling is increased by the multi-orbital renormalization, the density-induced tunneling is reduced. Coincidentally, both multi-orbital corrections compensate  each other  and the overall deviations can be approximated surprisingly well by the single-band bond-charge tunneling for moderate particle numbers (dashed line in \Fig{J}). This justifies the applicability of the model Hamiltonian \eqref{eq:Hsimple}, where tunneling is restricted to the lowest band, i.e., $J^\tot_{n_j,n_i} \approx J+(n_i+n_j-1)J_\BC$. 


\section{Occupation-number-dependent on-site interaction}
In a similar fashion as the inclusion of higher orbitals alters the tunneling, also the on-site interaction $U$ is modified. The orbital degree of freedom decreases the on-site interaction \cite{Will2010,Johnson2009,Buchler2010} as the particles tend to avoid each other. The results for $U^\MO_n$, representing the eigenvalue of the single-site exact diagonalization \cite{Will2010}, are depicted for occupation numbers $n=2$-$4$ in \Fig{Jtot}. 
Note that for strong interactions, it may be necessary to use realistic interaction potentials.
It is convenient to express $U^\MO_2$ within Eq.~\eqref{eq:Hsimple} by a fitted empirical function
	${\tilde U_2}/{ U}= \lambda_1+ \left(\lambda_2+\lambda_3 {\as}/{a}\right)\sqrt{{V_0}/{\ER}}
	+\lambda_4 e^{\lambda_5 \as/a} $,
which describes the on-site interaction for two particles but also serves as an approximation for three and four particles \cite{Note1}.

\section{Mean-field/Gutzwiller phase diagrams } \label{MFGW}
The SF-MI phase diagram (\Fig{PD}) for the extended Hamiltonian \eqref{eq:Htot} is derived within mean-field theory \cite{Oosten2001,Fisher1989}.  
By introducing the superfluid order parameter $\psi=\expect{\hatt b_i}=\expect{\hatt b_i^\dagger}$ and by neglecting the fluctuations of quadratic order the lattice sites are decoupled (see \ref{AppendixMeanField}).
Within second-order perturbation theory for the Mott state with $n$ particles, we find an analytical expression for the SF-MI transition point  
\begin{equation}
	\label{eq:SFMICriterion}
	\frac{n|\bar J_\BC (n-1) +1|^2 }{E_0(n)-E_0(n-1)} + \frac{(n+1)|\bar J_\BC\, n + 1|^2 }{E_0(n)-E_0(n+1)} + 1 =0
\end{equation}
with $\bar J_\BC=J^\BCMO_{n,n}/J^\MO_{n,n}$ and the unperturbed energy 
\begin{equation}
E_0(n)=(\frac12 U_n^\MO (n-1)-\mu) n/zJ_{n,n}^\MO.
\end{equation} 
Due to the decoupling in the mean-field approach, only tunneling processes with $n_i\!= \!n_j \!=\! n$ can be accounted for.
Thus, this approach cannot fully implement the occupation-number dependent Hamiltonian \eqref{eq:Htot}. This can be achieved by using the Gutzwiller approach, where the energy functional is minimized with respect to the coefficients $f_n$ of the trial wave function $\ket{G}= \prod_i \sum_n f_n \ket{n}_i$ (see \ref{AppendixGutzwiller}). The Gutzwiller results are depicted in \Fig{PD} (red lines) and show a nearly perfect agreement with the mean-field calculation (blue lines). 
 The mean-field solution \eqref{eq:SFMICriterion} can also be applied to the simplified occupation-independent model \eqref{eq:Hsimple} using the lowest-band tunneling parameters $J$ and $\bar J_\BC\!=\!J_\BC/J$ as well as the fitted on-site interaction $\tilde U_2$. 
Considering the drastic simplifications of this model, the predicted Mott lobes (green lines) are in compelling agreement with the results of the occupation-dependent Hamiltonian~\eqref{eq:Htot}.


\section{Conclusions}
We have presented an occupation-dependent model Hamiltonian incorporating multi-orbitally renormalized tunneling, density-induced tunneling and on-site interaction. The effectively increased tunneling and reduced on-site interaction cause a substantial modification of the Bose-Hubbard phase diagram for bosonic atoms. In addition, we have discussed a simplified occupation-independent model that includes the lowest-band bond-charge tunneling and reproduces the phase diagram well. In general, we have derived a multi-orbital renormalization procedure for one- and two-particle processes. It is capable of describing strongly-correlated optical lattice systems accurately that cannot be calculated correctly by means of variational mean-field treatments \cite{Li2006,Larson2009,Hazzard2010,Dutta2011}. 
Furthermore, it is important to include both bond-charge tunneling and multi-orbital corrections as extensions of the Hubbard Hamiltonian. Several effects such as the reduction of the bond-charge tunneling cannot be described with an effective wave function. The presented results can be transferred to fermionic systems, quantum gas mixtures, low-dimensional systems, long-range interactions and different lattice geometries. In these systems, the occupation-dependent tunneling may also lead to novel quantum phases. Related work on defining a basis with renormalized operators can be found in Ref.~\cite{Bissbort2011}.

\section{Acknowledgments} 
We thank Omjyoti Dutta, Maria Langbecker, Jannes Heinze, Maciej Lewenstein, Parvis Soltan-Panahi, and Ludwig Mathey for inspiring discussions. D.-S. L. acknowledges funding by the DFG grant GRK 1355.

\appendix 


\section{Multi-orbital interactions}
\label{AppendixModel}
The treatment of multi-band Hamiltonians for lattices represents a challenging  and still open problem posed by its complexity. A common approach is the restriction to the lowest band, which neglects interaction-induced admixtures of higher orbitals. In order to account for orbital degrees of freedom, we focus on two neighboring sites of the lattice, where we apply a multi-orbital diagonalization, so-called configuration interaction method, to each site individually (see Fig.~1d, main text). In this approach off-site interactions are neglected as the occupation of higher orbitals is mainly caused by the local on-site interaction. The three-dimensional Wannier functions 
\begin{equation}
	w_{\L/\R}^\a\ofr = w^\a(\mathbf{r}-\mathbf{r}_{\L/\R})
\end{equation}
constitute the single-particle orbitals, where 
	$\mathbf{r}_{\L/\R}=(\pm\frac12 a,0,0)$ 
is the coordinate of left and right site, respectively, $a$ the lattice spacing and 
\begin{equation}
	\alpha=(\alpha_x,\alpha_y,\alpha_z)
\end{equation}
the three-dimensional band index.
The maximally localized Wannier functions are obtained by means of band-structure calculations for a separable cubic lattice potential 	$V(\mathbf{r})=V_0\sum_i \cos^2(\pi r_i/a)$ with the lattice depth $V_0$.
To include correlations of particles fully, a many-particle Fock basis $\ket{N} = \ket{ n_0, n_1, ... }$ with $n$ particles is used, where $n_i$ is the number of particles in orbital $i$. 
The Hamiltonian for a single lattice site reads
\begin{equation}
	\label{eq:Hsite}
	\hat{H}_\text{site}=\sum_\alpha \epsilon^\a \hat{n}^\a + \frac{1}{2} \sum_{\alpha\beta\gamma\delta} U^\abcd\, \hbd{}{\alpha}\hbd{}{\beta}\hb{}{\gamma}\hb{}{\delta},
\end{equation}
where $\hat n^\a=\hbd{}{\alpha}\hb{}{\alpha}$, $\hbd{}{\alpha}$ creates and $\hb{}{\alpha}$ annihilates a particle in the Wannier orbital $\alpha$ with single-particle energies $\epsilon^\a$. 
In general, the ground state of the single-site problem with $n$ particles is given as a superposition of Fock states
\begin{equation} 
	\label{eq:ed}
	\ket{ \Psi(n) } = \sum_N c_N(n) \ket{ N(n) }
\end{equation} 
with real coefficients $c_N(n)$. Since the Hamiltonian \eqref{eq:Hsite} preserves parity the ground state consists of many-particle states with even parity. The lowest eigenvalue of the diagonalized Hamiltonian matrix directly corresponds to the occupation-dependent on-site interaction $U^\MO_n$ as discussed in the main text. 

We model the short-range interaction between the atoms as a box-shaped interaction potential
\begin{equation}
V(\mathbf{r}-\mathbf{r'})= \frac{\gt}{(2W)^3}  \prod_{i=1}^{3}  \theta \left(W-|r_i-r_i'|\right)
\end{equation}
with the Heaviside step function $\theta$ and a variable box width $W$. The influence of the finite interaction range depends only on the ratio $W/a$ and increases with the lattice depth since the wave functions are contracted. However, for $W \ll a$ the results (see \Fig{Jtot}, main text) are almost independent of $W$. The interaction integrals in Eq.~\eqref{eq:Hsite} are given by 
\begin{equation}
	U^\abcd(\gt, W)= \int\! d^3{r}\int\! d^3{r'}\ w^{\a *}\ofr\ w^{\b *}(\mathbf{r'})\ V(\mathbf{r}-\mathbf{r'})\ w^{(\gamma)}(\mathbf{r'})\ w^{(\delta)}\ofr,
\end{equation} 
where $\gt$ is fixed by the condition $U^{(0000)}(\gt, W)=g\dr |w\ofr|^4$ with $g=\frac{4\pi\hbar}{m}\as$ and s-wave scattering length $\as$.
Note that the $\delta$-interaction potential can be understood as the limiting case of the box interaction potential. However, it behaves differently if including an infinite number of orbitals requiring a regularization procedure as described in Ref.~\cite{Busch1998} (for numerical diagonalization see Ref.~\cite{Rontani2008}). For real interaction potentials, the mathematical subtlety of the $\delta$-interaction is not present and under certain experimental conditions \cite{Will2010} decay processes in higher orbitals may even effectively lead to a finite Hilbert space. 
Here, for the numerical diagonalization nine bands are accounted for in each spacial direction restricted to the energetically lowest $6 \times 10^3 \times n^2$ many-particle states.


\section{Multi-orbital tunneling}
\label{AppendixMOT}
\begin{figure}
\centering\includegraphics[width=0.5\linewidth]{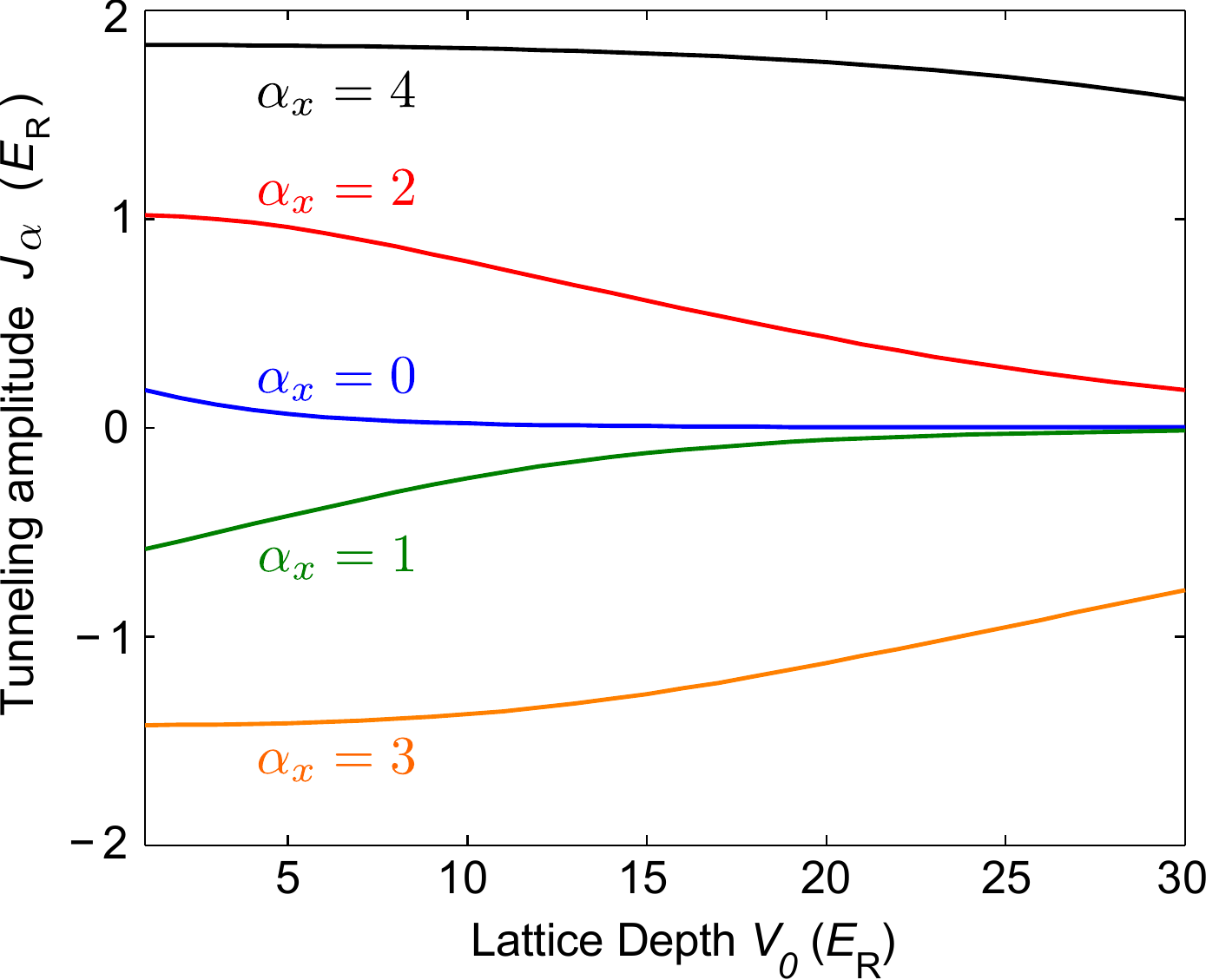}
\caption{Single-particle tunneling matrix elements $J_\alpha$ for the lowest five bands as a function of the lattice depth $V_0$ and in units of the recoil energy $\ER=h^2/8 m a^2$. \label{Jalpha}}
\end{figure}

In the following, we consider hopping processes, where one atom tunnels from the left site (L) with initially $\nL$ particles to the right site (R) with initially $\nR$ particles (see Fig.~1(d), main text). The tunneling operator accounting for all possible orbital hopping processes reads 
$	\hat{J}^\MO=\sum_{\alpha\beta} J_{\alpha\beta}\hbd{R}{\alpha}\hb{L}{\beta}$
with tunneling matrix elements
\begin{equation}
	\label{eq:J_ab}
	J_{\alpha\beta}=-\dr w^{\a *}_\mathrm{R}\ofr \left ( \frac{\mathbf{p}^2}{2m} + V\ofr \right) w^\b_\mathrm{L}\ofr.
\end{equation}
Because of orthogonality relations, only matrix element with $\alpha=\beta$ have non-zero contributions. Thus, the tunneling between left and right site simplifies to
\begin{equation}
	\hat{J}^{\MO} = \sum_\alpha J_\alpha\ \hbd{R}{\alpha} \hb{L}{\alpha}.
\end{equation}
with matrix elements $J_\alpha\equiv J_{\alpha\alpha}$, which only depend on the band index $\alpha_x$ in $x$ direction. The respective tunneling matrix elements $J_{\alpha}$ for the band $\alpha_x$ are plotted in \Fig{Jalpha}, which are positive for even bands and negative for odd bands. The tunneling amplitude is strongly enhanced for higher orbitals due to the considerably larger overlap of the Wannier functions. 

As the occupation of higher orbitals is decoupled from off-site interactions and tunneling in our model, it is possible to reduce the complex multi-band hopping to an effective hopping $J^\MO_{\nL,\nR}$. It describes the transition from an initial state $\ket{\Psi_\I}$ to the a final state $\ket{\Psi_\F}$ with 
\begin{equation}\begin{split}
	\label{eq:initialfinal}
	\ket{\Psi_{\I\;}}&=\ket{\Psi(\nL)}_\L\quad\;\,\ \otimes \ket{\Psi(\nR)}_\R , \\
	\ket{\Psi_\F}&=\ket{\Psi(\nL-1)}_\L\otimes \ket{\Psi(\nR+1)}_\R.
\end{split}\end{equation}
(see Fig.~1d). As discussed in the main text, the effective multi-orbital tunneling can be defined as
\begin{equation}
	\hat{J}^\MO_\mathrm{eff} = - \hatt{b}_\mathrm{R}^\dagger \hatt{b}_\mathrm{L} \, J^\MO_{\hatt n_\L,\hatt n_\R}
\end{equation}
with
\begin{equation}
	\label{eq:J_eff}
	J^\MO_{\nL,\nR}=\frac{\bra{\Psi_\mathrm{F}} {\hat{J}^\MO} \ket{\Psi_\mathrm{I}}}{\sqrt{\nL(\nR+1)}}.
\end{equation}
As inital and final states \eqref{eq:initialfinal} are product states, the expression separates into two terms for the left and the right site
\begin{equation}\begin{split}
\bra{\Psi_\mathrm{F}} {\hat{J}^\MO} \ket{\Psi_\mathrm{I}} = \sum_\alpha J_\alpha &\bra{\Psi(\nR+1)} \hbd{}{\alpha} \ket{\Psi(\nR)}_\R \\
 \times & \bra{\Psi(\nL-1) \!\ } \!\ \hb{}{\alpha} \; \ket{\Psi(\nL)}_\L .
\end{split}\end{equation}
Using the many-particle ground state \eqref{eq:ed} obtained by exact diagonalization, we can define
\begin{equation}\begin{split}
	\label{eq:j_a}
	j^\a_n=&\bra{\Psi(n-1)}\hb{}{\alpha}\ket{\Psi(n)}\\
  =&\sum_{N'N} c_{N'}(n-1) \ c_N(n) \bra{N'(n-1)}\, \hb{}{\alpha}  \ket{N(n)}.
\end{split}\end{equation}
Note that since the coefficients $c_N$ can be chosen to be real the conjugated term reads $\bra{\Psi(n-1)}\hbd{}{\alpha}\ket{\Psi(n)}={j}^{*\a}_{n-1}=j^\a_{n-1}$.
The effective tunneling can thus be written as
\begin{equation}
	\label{eq:JMO}
	J^\MO_{\nL,\nR}=\frac{1}{\sqrt{\nL(\nR+1)}}  \sum_\alpha J_\alpha \; j^\a_\nL \; j^\a_{\nR+1}.
\end{equation} 
Due to parity conservation of the Hamiltonian \eqref{eq:Hsite}, the ground states at the left and the right site are superpositions of Fock states $\ket{N(n)}$ with even parity. Tunneling in an odd orbital would alter the parity of both sites resulting in final states with vanishing coefficients. Consequently, tunneling can only occur in even orbitals $\alpha$ with positive tunneling matrix elements $J_\alpha$. It turns out that the effective tunneling is increased by the tunneling in higher bands with the exception of the processes $J^\MO_{n,0}=J^\MO_{1, n-1}$. Here, the lack of interaction in the left or the right site prohibits tunneling via higher bands. The interaction at one site, however, depopulates the Fock state $\ket{n,0,0,...}$ which results in slightly lowered tunneling energy. The process  $J^\MO_{1,0}$, where no interactions are present, is equivalent to the uncorrelated tunneling $J=J_0$.

Intuitively, the mean-field wave functions of the left and right site
\begin{equation}
w_\L^\mathrm{MF}\ofr = \sum_\alpha c^\a_{n_\L} w_\L^\a \ofr
\end{equation} and
\begin{equation}
w_\R^\mathrm{MF}\ofr = \sum_\alpha c^\a_{n_\R+1} w_\R^\a \ofr
\end{equation}
can be used to calculate the multi-orbital tunneling \cite{Dutta2011}.
The coefficients can be obtained from the orbital occupation numbers $c^\a_n=\sqrt{\expect{\hat n^\a_n}} / \sqrt{n} $, with the expectation value $\expect{\hat n^\a_n} = \braket{\Psi(n)|\hat n^\a|\Psi(n)}$. This allows to estimate the multi-orbital by
\begin{equation}
\begin{split}
J_{\nL,\nR}^\MO &\approx \dr w_\R^\mathrm{MF*}\ofr \left(\frac{\mathbf{p}^2}{2m}+V\ofr \right) w_\L^\mathrm{MF}\ofr \\
&= \sum_\alpha c^\a_\nL c^\a_{\nR+1} J_\alpha.
\end{split}
\end{equation}


\section{Off-site interactions}
\label{AppendixOffSite}
The full lowest-band interaction 
\begin{equation}   
	\hat H_\mathrm{int}= \frac{1}{2} \sum_{ijkl}  U_{ijkl}\,    \hat{b}^{\dagger}_i \hat{b}^{\dagger}_j \hat{b}_k \hat{b}_l 
\end{equation}
with interaction integrals \eqref{eq:Uijkl} is commonly restricted to the on-site interaction. However, off-site interactions between neighboring sites can reach the same order of magnitude as the tunneling.  Expanding the lowest-band Hamiltonian 
accordingly leads to three distinct physical processes, namely, bond-charge interaction (density-dependent hopping), correlated pair tunneling and density-density interaction \cite{Hirsch1989,Strack1993,Hirsch1994,Amadon1996,Mazzarella2006}. While the latter process has to be compared with the on-site interaction, the other two constitute tunneling processes. The full lowest-band Hamiltonian with nearest-neighbor interaction reads
\begin{equation}\begin{split}
 \hat H=	&- J \sum_{\langle i,j \rangle} \hat b_i^\dagger \hat b_j 
				- J_\BC \sum_{\langle i,j \rangle} \hat b_i^\dagger ( \hat n_i + \hat n_j ) \hat b_j 
				+ J_\text{pair} \sum_{\langle i,j \rangle}  \hat b_i^{\dagger 2} \hat b_j^{2} \\
				&+ \frac{1}{2} U \sum_i \hat n_i (\hat n_i -1)
				+ V \sum_{\langle i,j \rangle} \hat n_i \hat n_j 
\end{split}\end{equation}
with matrix elements $U=U_{iiii}$, $J_\BC=-U_{iiij}$, $J_\text{pair} = U_{iijj}/2$ and $V = U_{ijij}$. 

\begin{figure}
\centering\includegraphics[width=0.5\linewidth]{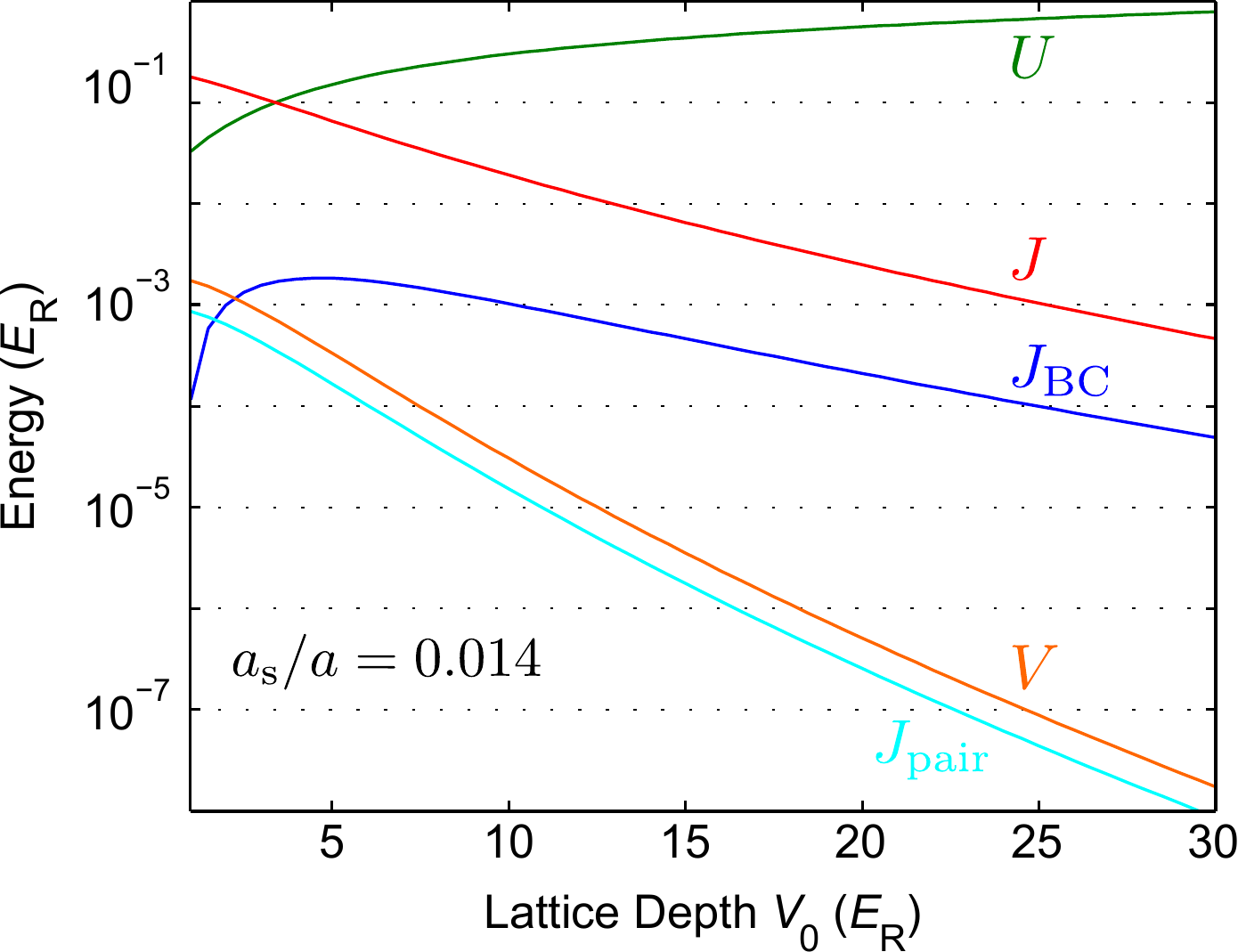}
\caption{Lowest-band parameters for on-site interaction $U$, tunneling $J$, bond-charge tunneling $J_\mathrm{BC}$, correlated pair-tunneling $J_\mathrm{pair}$ and density-density interaction $V$.\label{sbP}}
\end{figure}
In \Fig{sbP} these parameters are plotted as a function of the lattice depth. The off-site density-density interaction is very small compared to the on-site interaction and can consequently be neglected. This also applies for the correlated pair-tunneling, which is even negligible when compared with the single-particle tunneling matrix element $J$. The bond-charge tunneling matrix element, however, reaches ten percent of the conventional tunneling amplitude for intermediate and deep lattices. In addition, it scales with the total particle number and can thus be a significant contribution to the total tunneling. Note that the multi-orbital renormalization can influence the individual parameters very strongly.

\section{Multi-orbital bond-charge tunneling}
\label{AppendixBCT}
In analogy to the multi-orbital tunneling, the orbital degree of freedom also affects the bond-charge interaction. The single-band bond-charge or density-induced tunneling is introduced in the main text [see Eq.~(3) and Fig.~1(b)]. 
As discussed before, we can restrict the calculation to two neighboring sites and write the multi-orbital bond-charge operator (see Fig.~1(c), main text) as 
\begin{equation}
\begin{split}
   \label{eq:JBCMO}
    \hat{J}^\BCMO = \frac12 \sum_{\alpha\beta\gamma\delta} \hat{b}^{\a\dagger}_R  \Big(\  (&J^\BC_{\alpha\beta\gamma\delta} + J^\BC_{\alpha\beta\delta\gamma})\; \hat{b}^{\b\dagger}_L \hat{b}^{(\gamma)}_L  \\
								 + P \ (&J^\BC_{\delta\gamma\beta\alpha} + J^\BC_{\delta\gamma\alpha\beta})\; \hat{b}^{\b\dagger}_R\hat{b}^{(\gamma)}_R \ \Big)\  \hat{b}^{(\delta)}_L
\end{split}
 \end{equation}
 with
 \begin{equation}
\begin{split}
	J^\BC_{\alpha\beta\gamma\delta} = - &\int\! d^3{r}\int\! d^3{r'}w^\a_\R\ofr w^\b_\L(\mathbf{r'})\ V(\mathbf{r}-\mathbf{r'})\ w^{(\gamma)}_\L(\mathbf{r'}) w^{(\delta)}_\L\ofr \\
	= - P &\int\! d^3{r}\int\! d^3{r'}\  w^\a_\L\ofr w^\b_\R(\mathbf{r'})\ V(\mathbf{r}-\mathbf{r'})\ w^{(\gamma)}_\R(\mathbf{r'}) w^{(\delta)}_\R\ofr, 
\end{split}
\end{equation}
where the Wannier functions are chosen to be real and the sign $P=(-1)^{\alpha_x+\beta_x+\gamma_x+\delta_x}$ is parity-dependent. Note that for a $\delta$-interaction potential, we have $J^\BC_{\alpha\beta\gamma\delta} = J^\BC_{\alpha\beta\delta\gamma}$. The effective multi-orbital bond-charge tunneling reads
\begin{equation}
	\label{eq:JBC-eff}
	\hat{J}^\BCMO_\mathrm{eff}=-\hat{b}_\mathrm{R}^\dagger (\hat{n}_\R+\hat{n}_\L) \hat{b}_\mathrm{L} 
	\ J^\BCMO_{\hnL,\hnR}
\end{equation}
with occupation-number-dependent matrix elements
\begin{equation}
	\label{eq:JBCMO-parameter}
	J^\BCMO_{\nL,\nR} = \frac{\bra{\Psi_\mathrm{F}} \hat{J}^\BCMO \ket{\Psi_\mathrm{I}} }{ \sqrt{\nL(\nR+1)}(\nL + \nR - 1) }.
\end{equation}
Using that $\Psi_\mathrm{I}$ and $\Psi_\mathrm{F}$ are product states as well as the definitions \eqref{eq:j_a} and
\begin{equation}
	j^{(\beta\gamma\delta)}_n = \bra{\Psi(n-1)}\hbd{}{\beta} \hb{}{\gamma} \hb{}{\delta}\ket{\Psi(n)},
\end{equation}
the effective matrix elements can be computed by
\begin{equation}
	J^\BCMO_{\nL,\nR}=\sum_{\alpha\beta\gamma\delta} 
	\frac{ \frac12 (J^\BCMO_{\alpha\beta\gamma\delta} + J^\BCMO_{\alpha\beta\delta\gamma}) \left[j^\a_\nL j^{(\beta\gamma\delta)}_{\nR+1} + j^\a_{\nR+1} j^{(\beta\gamma\delta)}_\nL \right] }{ \sqrt{\nL(\nR+1)}(\nL + \nR - 1) }.
\end{equation}
Note that $\bra{\Psi(n-1)}\hbd{}{\beta} \hbd{}{\gamma} \hb{}{\delta}\ket{\Psi(n)} = j^{(\delta\gamma\beta)}_{n-1}$ and that because of the same arguments as in the last section, the orbitals $\alpha$ and $\beta+\gamma+\delta$ need to be even; thus the sign $P$ in equation \eqref{eq:JBCMO} vanishes.


\section{Effective potential analogy}
\label{AppendixEffectivePotential}
Considering bond-charge tunneling between neighboring sites $i$ and $j$ with the restriction to the lowest band, the bond-charge hopping simplifies to
\begin{equation}
   \label{eqA:JBC}
    \hat{J}_\BC = \hat{b}_i^\dagger\ (U_\mathrm{iiij}\,\hat{b}_i^\dagger\hat{b}_i \ 
		+ U_\mathrm{ijjj}\, \hat{b}_j^\dagger\hat{b}_j  )\ \hat{b}_j 
\end{equation}
(see Eq.~\eqref{eq:JBC}, main text), where $\hat{b}^{\dagger}_i=\hat{b}^{\0\dagger}_i $ and $\hat{b}_i=\hat{b}^\0_i$ are lowest-band creation and annihilation operators, respectively. The interaction integral reads
\begin{equation}
   \label{eqA:Uijkl}
 	U_{ijkl}=g  \dr
    w^{*}_i\ofr w^{*}_j\ofr w_k\ofr w_l\ofr
\end{equation}
with lowest-band Wannier functions 
	$w_i\ofr=w^\0(\mathbf r - \mathbf r_i)$. 
By inserting the integral expressions \eqref{eqA:Uijkl} in Eq.~\eqref{eqA:JBC} and using the commutation relations for  $\hat{b}^{\dagger}_i$ and $\hat{b}_i$, the bond-charge operator becomes
\begin{equation}\begin{split}
    \hat{J}_\BC = \hat{b}_i^\dagger \hat{b}_j \   
			g &  \left(   \dr  w^{*}_i\ofr \ \hat n_i |w_i\ofr|^2 \ w_j\ofr \right. 	\\ 
			& \; + \left. \dr  w^{*}_i\ofr\ (\hat n_j - 1 ) |w_j\ofr|^{2}\ w_j\ofr \right).
\end{split}\end{equation}
In order to combine bond-charge hopping and conventional tunneling $\hat J= - J\, \hat{b}_i^\dagger \hat{b}_j$ with
\begin{equation}
	\label{eq:J}
	J=-\dr w_i^*\ofr\left ( \frac{\mathbf{p}^2}{2m} + V\ofr \right) w_j\ofr, 
\end{equation}
we define a reduced density 
	$\hat\rho_{ij}=\hat n_i |w_i\ofr|^2 + (\hat n_j-1) |w_j\ofr|^2$ 	
on neighboring sites $i$ and $j$ excluding the hopping particle. Omitting the hopping particle corresponds to the exclusion of self-interactions. As a consequence, the total tunneling can be written as
 \begin{equation}\begin{split}
	\label{eqA:effectivePotential}
	\hat J^\BC_\eff& = \hat J + \hat{J}_\BC   \\
	 &=\! \dr w^{*}_i\ofr \!\left(\frac{\mathbf{p}^2}{2m}+ {V\ofr +g \hat\rho_{ij} }
	\right)\! w_j\ofr\,  \hat{b}^{\dagger}_i \hat{b}_j.
\end{split}\end{equation} 
In this integral, $\hat\rho_{ij}$ can be replaced  
by $\hat\rho_{ij}\to\rho\ofr -|w_j|^2$ with the density $\rho\ofr=\sum_i n_i |w_i\ofr|^2$ of all particles in the lattice, 
since the Wannier functions are localized. 
Finally, we can identify the effect of the bond-charge interaction as a normal tunneling within the effective potential
	$V_\mathrm{eff}= V\ofr +g(\rho\ofr -|w_j|^2)$
experienced by the hopping particle (see \Fig{J}c, main text). However, since $U_{iiij}=U_{ijjj}$, one may also define  
	$\hat\rho_{ij}=(\hat n_i-\frac12) |w_i|^2 + (\hat n_j-\frac12) |w_j|^2$.  In this case, the effective potential for a homogeneous filling $n\geq1$ can be written as $V\ofr + g \sum_i (n-\frac{1}{2})|w_i|^2$.
This form of the potential can be used to perform a band structure calculation in the effective potential \cite{Luhmann2008b}. Note, however, that in general the effective potential is not separable into its spatial directions. Furthermore, in a band structure calculus the Wannier functions are adapted to the effective potential whereas in Eq.~\eqref{eqA:effectivePotential} the Wannier functions of the bare lattice potential are used.       	


\section{Mean-field approach}
\label{AppendixMeanField}
The full Hamiltonian including the bond-charge interaction as well as the multi-orbital corrections to the tunneling, bond-charge and on-site interactions reads
\begin{equation}
\begin{split}
\label{eq:Hfull}
\hat H=&-\sum_{\expect{i,j}} \hatt{b}_i^\dagger \hatt{b}_j\, J^\MO_{\hatt{n}_j, \hatt{n}_i} -\sum_{\expect{i,j}} \hatt{b}_i^\dagger(\hatt{n}_i + \hatt{n}_j) \hatt{b}_j\, J^\BCMO_{\hatt{n}_j, \hatt{n}_i} \\
+ &\frac{1}{2}\sum_i U^\MO_{\hatt{n}_i} \hatt{n}_i(\hatt{n}_i - 1) -\mu \sum_i \hatt{n}_i
\end{split}
\end{equation}
(see Eq.~1, main text), where $\langle i,j \rangle$ denotes nearest-neighbor tunneling and $\mu$ is the chemical potential. The hopping in the renormalized multi-orbital basis is described by $\hatt{b}_i^\dagger \hatt{b}_j$, while $\hatt{n}_i$ corresponds to the number of particles on site $i$.
In order to apply mean-field theory (see Ref.~\cite{Oosten2001,Fisher1989}), we introduce the superfluid order parameter
	$\psi=\expect{\hatt b_i}=\expect{\hatt b_i^\dagger}$, where $\psi\neq0$ corresponds to the superfluid phase (SF) and $\psi=0$ defines the Mott insulator (MI) with a fixed number of atoms per lattice site.
The decoupling of the lattice sites is achieved by neglecting the fluctuations between $\hatt b_i^\dagger$ and $\hatt b_j$ of quadratic order, i.e.,  
\begin{equation}
\hatt{b}_i^\dagger \hatt{b}_j \approx \hatt{b}_i^\dagger \hatt{b}_j - (\hatt{b}_i^\dagger - \braket{\hatt{b}_i^\dagger})(\hatt{b}_j - \braket{\hatt{b}_j}) =  \psi (\hatt{b}_i^\dagger + \hatt{b}_j) - \psi^2.
\end{equation}
However, the occupation-number dependent tunneling parameter still couples the sites and it is necessary to approximate
\begin{equation}
 \sum_{\expect{i,j}}(\psi (\hatt{b}_i^\dagger + \hatt{b}_j) - \psi^2) J^\MO_{\hatt{n}_j, \hatt{n}_i} \approx z \sum_i (\psi (\hatt{b}_i^\dagger + \hatt{b}_i) - \psi^2) J^\MO_{\hatt{n}_i, \hatt{n}_i},
\end{equation}
which corresponds to substituting $J^\tot_{n_i\pm1,n_i}$ and $J^\tot_{n_i,n_i\pm1}$ by $J^\tot_{n_i,n_i}$. This is necessary, as the fluctuations are described by a single parameter $\psi$, which has no internal structure that would allow for the implementation of an occupation-dependent tunneling amplitude.
This can be circumvented by using the Gutzwiller approach, where the information on the population of the individual Fock states is preserved (see \ref{AppendixGutzwiller}). 
However, we find that this approximation causes only small deviations.
Analogously, we neglect quadratic fluctuations in the bond-charge term between neighboring lattices sites. Disregarding terms of the order $\mathcal{O}(\psi^3)$ we find
 \begin{equation}
	\hatt{b}_i^\dagger \hatt{n}_i \hatt{b}_j + \hatt{b}_i^\dagger \hatt{n}_j \hatt{b}_j 
	\approx \psi(\hatt{b}_i^\dagger \hatt{n}_i + \hatt{n}_j \hatt{b}_j).
\end{equation}
We perform second-order perturbation theory in $\psi$ of a Mott lobe with $n$ particles per site and thus restrict the tunneling to the symmetric terms $J^\MO_{n, n}$ and $J^\BCMO_{n, n}$.
This results in a decoupled single-site Hamiltonian $\hat H_i^\mathrm{eff} / z J^\MO_{n, n} = \hat H_0 + \psi \hat V$ with
\begin{equation}
\begin{split}
	\hat H_0 &= \frac{1}{2} \bar{U}_{\hat n} \hat{n}(\hat{n}-1) - \bar{\mu}\hat{n} + \psi^2, \\
	\hat V &= -(\hat{b}^\dagger + \hat{b}) - (\hat{b}^\dagger \hat{n} + \hat{n} \hat{b}) \bar J^\BC_{n,n},
\end{split}
\end{equation}
 $z=6$ the number of nearest neighbors, and  $(\bar U_{\hat n}, \bar J^\BC_{n,n},\bar\mu) = (U_{\hat n}^\MO, J_{n,n}^\BCMO,\mu)/zJ_{n, n}^\MO$. The unperturbed energy for a Mott state  $\ket{n}=\frac{1}{\sqrt{n!}} \hat b^{\dagger n} \ket{0} $ with $n$ particles is given by 
\begin{equation}
	E_0(n)=\bra{n} \hat H_0\ket{n}=\frac{1}{2}\bar{U}_n n(n-1) - \bar{\mu} n .
\end{equation}
The perturbation series up to third order in $\psi$ reads
\begin{equation}
	E(\Psi)=E_0(n) + E_2(n)\Psi^2 + \mathcal{O}(\psi^4) \\
\end{equation} 
with the second-order correction
\begin{equation*}
	E_2(n)=\frac{n|\bar{J}^\BC_{n,n}(n-1)+1|^2}{E_0(n)-E_0(n-1)} + \frac{(n+1)|\bar{J}^\BC_{n,n}\ n+1|^2}{E_0(n)-E_0(n+1)} 
	+ 1.
\end{equation*} 
The boundary of the  SF-MI phase transition is given by the Landau criterion $E_2(n)=0$ for second-order phase transitions, which can be solved for $\mu$ (see Fig.~1, main text).


\section{Gutzwiller approach}
\label{AppendixGutzwiller}
In the mean-field approach elaborated above, only the tunneling matrix elements $J^\MO_{n,n}$ and  $J^\BCMO_{n,n}$ are taken into account due to the decoupling approximation. Using  a Gutzwiller approach \cite{Krauth1992,Rokhsar1991}, the full occupation-dependent Hamiltonian \eqref{eq:Hfull} can be evaluated without this restriction. In this approach, the trial wave function
\begin{equation}
\label{eq:GW_trial}
	\ket{G_i}=\sum_{n=0}^\infty f_n \ket{n}_i 
\end{equation}
with coefficients $f_n$ is used, which assumes that the wave function at each site $i$  can be written as a sum of local Fock states. In a homogeneous lattice with $M$ sites, all sites are equivalent and we can write $\ket{G}=\prod_{i=1}^M \ket{G_i}$.
The Hamiltonian \eqref{eq:Hfull} can be split into four parts, namely,  
\begin{equation}
	\hat{H} = \hat{J}_\mathrm{eff} + \hat{J}^\BC_\mathrm{eff} + \hat{U} + \hat{\mu}. 
\end{equation}
The expectation values for this Hamiltonian, depending exclusively on the expectation values for two adjacent sites $i$ and $j$, are given by
\begin{equation}
\begin{split}
\braket{\hat{J}_\mathrm{eff}} =& -zM \bra{G_i G_j} \hatt{b}^\dagger_i \hatt{b}_j J^\MO_{\hatt{n}_j, \hatt{n}_i} \ket{G_i G_j} \\
=& -zM \sum_{n_i,n_j} f_{n_i+1} f_{n_i} f_{n_j +1} f_{n_j} \\
&\times \sqrt{n_i+1}\sqrt{n_j+1} J^\MO_{n_j+1, n_i},
\end{split}
\end{equation}
\begin{equation}
\begin{split}
\braket{\hat{J}_\mathrm{eff}^\BC} =& -zM  \bra{G_i G_j} \hatt{b}^\dagger_i (\hatt{n}_i+\hatt{n}_j) \hatt{b}_j J^\BCMO_{\hatt{n}_j, \hatt{n}_i} \ket{G_i G_j} \! \\
=& -zM \sum_{n_i,n_j} f_{n_i+1} f_{n_i} f_{n_j +1} f_{n_j} \\
& \times (n_i+n_j)\sqrt{n_i+1}\sqrt{n_j+1} J^\BCMO_{n_j+1, n_i},
\end{split}
\end{equation}
\begin{equation}
\braket{\hat{U}} = \frac{1}{2}M \sum_n U_n n(n-1) |f_n|^2,
\end{equation}
\begin{equation}
\braket{\hat{\mu}}= - M \sum_n \mu n |f_n|^2,
\end{equation}
with $z=6$ nearest neighbors.
The system is in a Mott-insulator state, when all fluctuations vanishes, i.e., when $\ket{G}=\prod_i\ket{n}_i$ is the energetically lowest state. The resulting Gutzwiller phase diagram is plotted together with the mean-field results in Fig.~1 in the main text. Both approaches, Gutzwiller and mean-field, agree almost perfectly.


\end{document}